\documentclass[twocolumn]{emulateapj}
\usepackage{amsmath}
\usepackage{rotating}
\usepackage{hyperref}
\usepackage{epsfig}

\begin{document}
\def\gapprox{\mathrel{\vcenter{\offinterlineskip \hbox{$>$}
    \kern 0.3ex \hbox{$\sim$}}}}
\def\lapprox{\mathrel{\vcenter{\offinterlineskip \hbox{$<$}
    \kern 0.3ex \hbox{$\sim$}}}}
\newcommand{\beq}{\begin{equation}}
\newcommand{\eeq}{\end{equation}}

\def\eps{\epsilon}
\def\epsrad{\eps_{\rm rad}}
\def\epsw{\eps_{\rm w}}
\def\epsj{\eps_{\rm jet}}

\def\lnu{L_{\nu}}
\def\Edotw{\dot{E}_{\rm w}}
\def\Lbh{L_{\rm BH}}
\def\Ledd{L_{\rm Edd}}
\def\Ewind{E_{\rm wind}}
\def\ergs{\rm erg~s^{-1}}

\def\Msun{M_{\odot}}
\def\Mbh{M_{\rm BH}}
\def\Medd{M_{\rm Edd}}
\def\Macc{M_{\rm acc}}
\def\Minf{M_{\rm inf}}
\def\Moutf{M_{\rm outf}}
\def\Mwind{M_{\rm wind}}

\def\Mdotacc{\dot{M}_{\rm acc}}
\def\Mdotinf{\dot{M}_{\rm inf}}
\def\Mdotoutf{\dot{M}_{\rm outf}}
\def\Mdotbh{\dot{M}_{\rm BH}}

\def\vw{v_{\rm w}}
\def\vwten{v_{\rm w,10}}
\def\kms{\rm km~s^{-1}}
\def\pdot{\dot{p}}

\def\Re{R_{\rm e}}
\def\Rbondi{r_{\rm Bondi}}

\shortauthors{Choi et al.}
\shorttitle{Radiative and Mechanical AGN Feedback}

\title{Radiative and Momentum Based Mechanical AGN Feedback 
                                 in a 3-Dimensional Galaxy Evolution Code}

\author{Ena Choi\altaffilmark{1}, Jeremiah P. Ostriker\altaffilmark{1,2},
Thorsten Naab\altaffilmark{3} and Peter H. Johansson\altaffilmark{4,5}}
\affil{$^1$Department of Astrophysical Sciences, Princeton University, 
Princeton, NJ 08544, USA}
\affil{$^2$IoA, Cambridge, UK}
\affil{$^3$ Max-Planck-Institut f\"ur
  Astrophysik, Karl-Schwarzschild-Strasse 1, 85741 Garching, Germany} 
\affil{$^4$ Department of Physics, University of Helsinki, 
Gustaf H\"allstr\"omin katu 2a, FI-00014 Helsinki, Finland}
\affil{$^5$ Finnish Centre for  Astronomy with ESO, University of Turku, 
V\"ais\"al\"antie 20, FI-21500 Piikki\"o, Finland}

\begin{abstract}
We study the growth of black holes (BHs) in galaxies using 
three-dimensional smoothed particle hydrodynamic simulations 
with new implementations of the momentum mechanical feedback, 
and restriction of accreted elements to those that are gravitationally 
bound to the BH. We also include the feedback from the X-ray radiation 
emitted by the BH, which heats the surrounding gas in the host 
galaxies, and adds radial momentum to the fluid. We perform simulations 
of isolated galaxies and merging galaxies and test various feedback 
models with the new treatment of the Bondi radius criterion. We find that 
overall the BH growth is similar to what has been obtained by 
earlier workers using the Springel, Di Matteo, \& Hernquist algorithms. 
However, the outflowing wind velocities and mechanical energy emitted 
by winds are considerably higher ($\vw \sim 1000-3000~\kms$) 
compared to the standard thermal feedback model ($\vw \sim 50-100~\kms$). 
While the thermal feedback model emits only 0.1\% of BH 
released energy in winds, the momentum feedback model emits more 
than 30\% of the total energy released by the BH in winds. In the 
momentum feedback model, the degree of fluctuation in both radiant and 
wind output is considerably larger than in the standard treatments. We 
check that the new model of the BH mass accretion agrees with analytic 
results for the standard Bondi problem.

\end{abstract}

\section{Introduction}\label{intro}
Accretion of matter onto a supermassive black hole (SMBH) is believed 
to emit enormous amounts of electromagnetic, luminous radiation and 
drive powerful jets, winds or outflows of quasar \citep{lyn69,ree84}. The 
energy outputs emitted by accreting SMBHs at the centers of bulges 
and elliptical galaxies \citep{kor95,ric98} are believed to play an 
important role during galaxy formation and evolution, as revealed by 
many empirical correlations between their masses $\Mbh$ and host 
galaxy properties, e.g., the dispersion $\sigma$ of the host galaxy 
\citep{geb00,tre02,gul09,gra11}, bulge stellar mass \citep{dre89,kor93,
mag98,mar03,gul09}, and the bulge binding energy \citep{all07,bar07}.  
These correlations have led to the development of models where the 
SMBHs are linked by feedback from the central SMBHs, i.e., 
feedback via the mass ejection by winds or jets, or the emitted radiation 
regulates the mass accretion rate and the final SMBH mass. The 
feedback can be in the forms of radiatively or mechanically driven winds 
\citep{fab99,pro00,pro08}, or of radiative effects such as Compton 
heating and photoionization \citep{saz05,cio07}, or of radiation pressure 
\citep{deb10,deb10b}. Thermal feedback, heating by an unspecified 
mechanism has also been employed by many authors, e.g., 
\citet{spr05b,hop05,joh09b,joh09}.

In this rapidly developing field of active galactic nucleus (AGN) feedback,
it is natural for large scale numerical simulations to explore first the 
simplest schemes. Thus physical processes, which cannot all be 
included in the first three-dimensional hydrodynamic simulations, have 
been treated in a selective fashion. For example, in \citet[here after 
``SdMH05'']{spr05b}, among the first who reproduced the AGN feedback
in a three-dimensional smoothed particle hydrodynamics (SPH) code, 
the feedback was assumed to be purely thermal.   That is, some 
fraction of the bolometric luminosity of accreting BHs is deposited as 
thermal energy to the neighboring gas particles via a mechanism that is 
not specified. The authors found that this thermal feedback treatment 
regulates and then terminates the further growth of the BH and expels 
gas from the central region in the galaxy in a quasar-driven outburst. 
This pioneering work, however, did not specify how the injected energy 
reaches the thermally heated gas particles. The conveyance of the 
energy is likely to be via either radiation or a wind, and in both cases 
momentum must be added with the thermal energy. If the energy is 
transferred via a wind with velocity $\vw$, then the mass transfer may be 
significant and, since the ratio of momentum to energy scales as $1/\vw$, 
the momentum transfer is correspondingly larger.

In a first attempt to study the relative importance of the different 
processes, utilizing one- and two-dimensional computations, it was 
found that momentum injection is the dominant mode of feedback
\citep[here after ``OCCNP10'']{ost10}, because the very short 
cooling time of dense gas makes thermal input relatively inefficient. 
At the high densities of the central region of the galaxies, the cooling 
time of gas is sufficiently short in high resolution simulations, so that 
it cannot retain the injected thermal energy and efficiently convert it to 
kinetic energy. However, if the input to the surrounding fluid is via 
winds, then the return of mass and momentum to that fluid can be the 
dominant drivers which can reduce the accreted SMBH mass by up to a 
factor of 100 (OCCNP10). Other recent work by \citet{deb10,deb10b}, 
emphasizing the momentum input from optically thick radiation fields 
($\sim \tau L/c$ with $\tau \sim 10$) has also found the dramatic effects
of momentum input.

Observationally, around 15\%--20\% of bright quasars 
\citep{hew03,dai08,gib09}  show outflows, where blueshifted broad 
absorption lines (BAL) are attributed to subrelativistic ($\sim10,000~\kms$)
mass ejection. According to the fraction of quasars with BAL 
winds together with the assumption that all quasars have outflows, the 
BAL outflows are wide, covering at least 20\% of the solid angle. The 
radiation-driven winds from the detailed hydro simulation by \citet{pro04} 
cover $\sim \pi$ str, supporting this observational estimate.
These outflows inject mass, momentum, and energy into the 
surrounding gas and are believed to be more efficient feedback 
agents of the host galaxies than relativistic jets, which drill through the 
ambient gas and put most of the energy into the intergalactic medium. 

However, the effects of outflows to the host galaxies and larger 
cosmological structures are just beginning to be studied, because their 
mass ejection rates, energetics and sizes were largely undetermined. 
Adopting the recent measurements of mass ejection rates, and kinetic 
luminosities of the outflows from the absorption-line observations 
\citep{ara08,moe09,dun10}, many authors introduced the kinetic AGN 
feedback in adaptive mesh refinement (AMR) simulations 
\citep{omm04, kim11,dub09} and  in SPH simulations \citep{nay10,
deb11}.

It has been shown that the average spectrum of AGN has a strong 
secondary peak in the high-energy X-rays ($\sim$ 50--100 keV) which is 
the main contributor to the Compton and secondary metal line heating 
\citep{saz04}. Some recent studies using one- and two-dimensional 
simulations including \citet{cio07,nov11} considered the effect of 
momentum input, heating and radiation pressure from the AGN radiation.
The recent SPH work by \citet{ham11} found that X-ray feedback, the 
heating and radiation pressure associated with the X-ray radiation field 
emitted from the BH, is more effective at suppressing star formation and
BH mass growth compared to the traditional thermal feedback approach.
The X-ray heating was also found to be effective in AMR simulations 
\citep{kim11}, as it heats the surrounding interstellar medium (ISM), keeps it 
hot for an extended duration of time, and effectively self-regulates the 
growth of the BH. 

In many previous numerical studies of galaxy formation and evolution 
that only can resolve hundreds of pc to kpc scales, the accretion of gas 
onto BHs at the centers of AGNs occurs on unresolved scales. The 
Bondi--Hoyle--Lyttleton formalism \citep{hoy39,bon44,bon52} is 
commonly adopted, to obtain the BH mass accretion rate from the 
resolved larger scale properties of the ambient gas. This assumption 
has been made in studies of isolated galaxy and merging galaxies 
\citep[e.g., SdMH05,][]{you08,joh09} and as well in cosmological
simulations \citep[e.g.,][]{sij07,dim08,boo09,tey11,dub11}. 

However, in most of the cases treated to date, the SPH smoothing length 
or the numerical resolution of the respective code is larger than the 
Bondi radius, i.e., the inner flow is numerically unresolved. The 
limitation of numerical and spatial resolution may introduce the unusual 
and unwanted effect---the neighboring gas particles around a BH at any 
distance from the BH can be accreted even though they are not 
gravitationally bound to the BH. This physically awkward treatment of the
BH mass accretion may result in the very large growth of the central BH if 
we used the standard accretion algorithm without feedback. We will 
introduce a new treatment of gas particle accretion, ``Bondi radius 
criterion'',  which statistically limits the accretion of mass to the gas 
particles which are within the Bondi radius.

The purpose of the current paper is to introduce (1) a modeling of AGN 
mechanical feedback via winds as observed in BAL systems that 
includes mass and momentum feedback, as well as thermal input; (2) 
the detailed treatment of radiative effects, i.e., X-ray radiative 
feedback; and (3) a modified BH accretion rate prescription using the 
Bondi radius criterion in the parallel TreeSPH-code GADGET. In this 
paper, we restrict our exploration to the simulations of isolated disk 
galaxies and merging galaxies, in order to better understand the 
specific properties of the new treatments. We reserve the follow-up
papers for detailed analysis of merger simulations and cosmological
simulations, to be compared with the results of observational papers.

This paper is structured as follows. In Section~\ref{model}, we describe 
the simulation code, and  discuss the BH feedback and new BH 
accretion prescription. We present the results and comparisons 
between our new model and standard prescription in 
Section~\ref{result}. We summarize and discuss our findings in 
Section~\ref{discussion}.

\section{The Models}\label{model}
\subsection{Numerical Code}
We perform the simulations using the parallel TreeSPH-code GADGET
\citep{spr05}. The code employs the Lagrangian SPH \citep[see][]{mon92} 
technique and solves the equations of motion for the collisionless dark 
matter and star particles and gas. We include the radiative cooling for a 
primordial mixture of hydrogen and helium \citep{kat96} and a spatially 
uniform time-independent local UV background \citep{haa96}. The gas 
of the ISM is assumed to be a two-phase medium of hot 
and cold gas \citep{mck77,spr03} and stars form from a cold component
embedded in sufficiently dense gas, i.e., $\rho > \rho_{\rm th}$ with 
the short-lived stars supplying an energy of $10^{51}$~erg to the 
surrounding gas per supernovae (SNe). This SN feedback heats the hot 
phase of the ISM and evaporates cold clouds, establishing a 
self-regulation cycle for star formation. SN-driven galactic winds are not 
included in this study.

We include all the basic aspects of the model for black hole (BH) 
accretion and feedback adopted in SdMH05, \citet{spr05a,joh09}, and 
implement the momentum and mass feedback. 

\subsection{Black Hole Feedback Model}\label{sec:bhfeedback}
In the widely adopted SdMH05 model, the numerically unresolved 
accretion onto the BH is related to the large scale resolved gas 
distribution using a Bondi-Hoyle-Lyttleton parameterization 
\citep{hoy39,bon44,bon52}. In very high resolution treatments where 
the Bondi radius ($R_B \sim 2GM_{\rm BH}/(c_s^2 + v^2)^{1/2}$) is 
resolved such as  \citet{cio10,nov11,bar11}, there is no need to specify 
the accretion rate as the hydro code with an appropriate inner 
boundary condition will correctly calculate the accretion rates. The 
accretion rate onto the BH in unresolved flows is estimated as
\begin{equation}
\dot{M}_{\rm{B}}=\frac{4 \pi \alpha G^{2} M_{\rm BH}^{2} \rho}
                            {(c_{\rm s}^2+v^{2})^{3/2}},
\label{Bondi}
\end{equation}
where $\rho$ and $c_{\rm s}$ are the density and sound speed of the
surrounding gas, respectively. $v$ is the velocity of the BH relative to the
surrounding gas. Here $\alpha$ is a dimensionless parameter setting 
the efficiency of the accretion and it is conventionally set to be 
$\alpha=100$ in SPH simulations on the grounds that the low spatial 
resolution currently available would otherwise limit the accretion rate to 
lower than the true value. Adopting $\alpha=100$ gives reasonable 
results for the low resolution simulations discussed in SdMH05, and 
\citet{joh09}, but in general $\alpha$ is resolution dependent. We adopt 
different values of $\alpha$ for the different resolutions, and details will 
be presented later in Section~\ref{simulations}. Note that 
Equation~(\ref{Bondi})  describes the accretion onto a point mass 
surrounded by adiabatic ($\gamma=5/3$) gas with properties $\rho$, 
$c$, and $v$ (in Equation~(\ref{Bondi})) far away from the BH 
($r \rightarrow \infty$). Usually it is also assumed that the maximum 
accretion is limited to the Eddington rate given by
\begin{equation}
\dot{M}_{\rm{edd}} \equiv \frac{4 \pi G M_{\rm BH} m_{\rm{p}}}
                                                  {\epsilon_{\rm{r}}\sigma_{\rm{T}} c}.
\label{Eddington}
\end{equation}
Here $m_{\rm{p}}$ is the proton mass,  $\sigma_{\rm{T}}$ is the 
Thomson cross-section and $\epsilon_{\rm{r}}$ is the radiative efficiency
assumed to be a fixed value of 0.1 adopted from the mean value for 
radiatively efficient \cite{sha73} accretion onto a Schwarzschild BH.
This value is also supported by the global mass--energy relation
pointed out by \citet{sol82} and \citet{yu02}. The accretion rate is then
$\dot{M}_{\rm{inf}}=\rm{min}(\dot{M}_{\rm{B}},\dot{M}_{\rm{edd}})$.
The BHs are represented by collisionless ``sink'' particles, which
feel only gravitational forces. The properties, including density, 
temperature, and the bulk velocity of the local gas around the BH, which 
then define the accretion rate, are estimated in a similar fashion to 
normal SPH particles.

Technically, the actual accretion of gas particles onto the BH particle is 
implemented using a stochastic approach (SdMH05). For each gas 
particle $j$ around a BH,  the probability of being absorbed by the BH 
is calculated as
\beq
 p_{j} = \frac{w_j \dot{M}_{\rm{inf}}  \Delta t} {\rho}  ,
\label{p_acc}
\eeq
where $w_j$ is the kernel weight of the gas particle relative to the BH,
$\dot{M}_{\rm{inf}}$ is the BH accretion rate and, $\Delta t$ is the time step.
The gas density $\rho$ is measured at the position of the BH as
\beq
\rho=\sum\limits_{i=1}^N m_i w_i,
\label{eq:rho}
\eeq
where $m_i$ denotes the gas particle mass, and $N$ is the number of 
neighboring particles to the BH. From Equations~(\ref{p_acc}) and 
(\ref{eq:rho}), we see that the probability of the $j$th particle being 
accreted in time interval $\Delta t$ is close to 
$ p_j = \dot{M}_{\rm{inf}} \Delta t / m_{\rm gas}$, but influenced by the 
smoothing kernel of gas particle near the BH. The gas particle is 
swallowed by the BH when the probability $ p_j$ is larger than the 
generated random number uniformly distributed in the interval [0,1].

\subsubsection{Thermal Feedback}
In previous GADGET based studies, the feedback energy from the BH
${E}_{\rm feed}$ has typically been assumed to be some fraction 
$\epsilon_{\rm{f}}$ of the radiated luminosity $L_{\rm{r}}$ and couples 
thermally and isotropically to the surrounding gas as,
\begin{equation}
\dot{E}_{\rm feed}=\epsilon_{f} L_{\rm{r}}= 
                               \epsilon_{f} \epsilon_{\rm{r}} \dot{M}_{\rm{inf}}c^{2}.
\end{equation}
Many authors (e.g., SdMH05) adopt a fixed value of 
$\epsilon_{f}=0.05$ so that $\epsw=\epsilon_{f} \epsilon_{\rm{r}}=
5\times10^{-3}$, i.e.,  0.5 percent of the accreted rest mass energy in 
total is available as thermal energy feedback. Utilizing this value for 
thermal feedback and adopting $\alpha=100$, the normalization of the 
$M_{\rm BH}-\sigma$ was recovered by \citet{dim05} in disk mergers 
simulations having a spatial resolution of $\epsilon_{\rm gas}=0.1 h^{-1}$ 
kpc and a mass resolution of $M_{\rm gas} \sim 6 \times 10^5 \Msun$.

The corresponding feedback energy is distributed as thermal energy to 
the surrounding $\sim 64$ gas particles weighted by the SPH kernel. 
Note that the results may depend on the details of the numerical 
implementation.  If $\sim 128$ particles are used of a given mass rather 
than $\sim 64$, the sound speed $c_{\rm s}^2$ would have been lower 
with corresponding increase of the accretion rate $\dot{M}_{\rm{B}}$ in 
Equation~(\ref{Bondi}). What matters is the mass of the material into which
the thermal energy is dumped. Thus, if the mass per particle were halved
and the particle number was doubled the results would not change. In 
this standard approach, neither mass nor momentum is added to the 
ambient fluid by the BH and all energy that is added is via 
thermal energy.

\subsubsection{Momentum Feedback}\label{sec:momentum}
Accreting BHs are observed to emit broad-line winds that convey mass, 
energy, and momentum into the surrounding gas \citep{dek01,moe09,
dun10} and our goal is to include these observed flows in our numerical 
treatment. To implement mechanical momentum and mass feedback, 
we define the inflowing and outflowing mass rates to be ($\Mdotinf$, 
$\Mdotoutf$),  and we use the following simple equations based on the 
conservation of mass, energy, and momentum (cf. OCCNP10):
\beq
\Mdotacc = \Mdotinf - \Mdotoutf, \label{eq:Mdot}
\eeq
where $\Mdotacc$ is the mass rate actually accreted onto the BH. We 
define the kinetic energy rate of the outflow $\Edotw$ as,
\begin{subequations}
    \begin{eqnarray}
\Edotw & \equiv & \epsw \Mdotacc c^2, \label{eq:edotw1} \\
      &=& \frac{1}{2} \Mdotoutf \vw^2,\label{eq:edotw2} 
    \end{eqnarray}
\end{subequations}
\beq
\pdot = \Mdotoutf \vw, \label{eq:pdot}
\eeq
where we have oversimplified matters by allowing only one wind 
velocity, $\vw$, when in fact Equation~(\ref{eq:edotw2}) requires 
$\langle \vw^2 \rangle$ and Equation~(\ref{eq:pdot}) requires 
$\langle \vw \rangle$ and $\epsw$ denotes the feedback efficiency, 
i.e., the wind efficiency in the momentum feedback model. Now, 
defining
\beq
\psi \equiv 2 \epsw c^2 / \vw^2=\Mdotoutf / \Mdotacc,
\label{eq:psi}
\eeq
we have, as solutions to Equations~(\ref{eq:Mdot}) \-- (\ref{eq:pdot}),
\begin{subequations}
    \begin{eqnarray}
\Mdotacc &=& \Mdotinf \frac{1}{1+\psi}\label{eq:Mdot_sol} ,\\
\Mdotoutf &=& \Mdotinf\frac{\psi}{1+\psi}, \\
\Edotw     &=& \epsw c^2 \Mdotinf \frac{1}{1+\psi}, \\
\pdot        &=& \Mdotinf\vw \frac{\psi}{1+\psi}.
    \end{eqnarray} \label{eq:sol}
\end{subequations}

As we see from Equations (\ref{eq:sol}a) and (\ref{eq:sol}b), there is an important 
dimensionless quantity $\psi \equiv \Mdotoutf / \Mdotacc$. In typical 
treatments of AGN feedback, $\psi$ is assumed to be $0$ in equations 
(\ref{eq:sol}) implicitly assuming $\vw \to \infty$, and so $\Mdotoutf$ and 
$\pdot$ are neglected. Thus, the two terms $\Edotw$ and $\Mdotacc$ 
may be overestimated, and $\pdot$, the momentum input to the 
surrounding fluid is neglected. For example, if we adopt for the efficiency 
of generating mechanical energy the value $\epsw = 
\epsilon_{f} \epsilon_{\rm{r}} =5 \times 10^{-3}$, as adopted by 
\citet{spr05} and other authors, and we take $\vw = 10,000~\kms$
($\vwten$) \citep{cre03,moe09}, then we have $\psi = 9~ \vwten^{-2}$ 
and all of the neglected effects may in fact be dominant; 90 \%, 
($\psi/(1+\psi)$) of the inflowing mass may be ejected in a disk broad 
absorption line (BAL) wind and the mass and momentum input 
deposited in the ambient gas may dominate over the energy input.

To implement the actual output of mass, momentum, and energy, we 
modify the stochastic approach applied for the gas particle accretion on 
the BH shown in Equation~(\ref{p_acc}). We first calculate a probability of 
being attracted into the central zone by the BH for each gas particle 
nearby using Equation~(\ref{p_acc}) and determine its fate by generating 
a random number $x_j$ in the interval [0,1]. For $x_j < p_j$, the gas 
particle is taken to be part of the inflow onto the BH. We then draw an 
independent random number $y_j$ uniformly distributed in the interval 
[0,1] and compare it with $1/(1+\psi)$, the probability of being actually 
absorbed by the BH. For $y_j < 1/(1+\psi)$, the gas particle is accreted 
onto the BH. For $y_j > 1/(1+\psi)$, the gas particle is ejected with its 
wind velocity $\vw$ and momentum.

We fix the wind velocity $\vw$ to 10,000 $\kms$ ($\vwten$) 
\citep{cre03,moe09} corresponding to a typical broad line wind. 
Together with our choice of $\epsw=5 \times 10^{-4}$ (note that our 
energy coupling parameter $\epsw$ is 10 times smaller than the value 
used in other GADGET-2-based simulations), we have $\psi=0.9$. That
is, essentially one of the two particles inflowing to the BH is actually 
accreted by it, the other is driven out as part of the broad absorption line 
wind.

We emit a particle radially from the BH in the specified direction. We set 
the direction of wind to be parallel or anti-parallel to the direction of 
angular momentum ($\vec{r} \times \vec{v}$) of each gas particle, 
making it to be essentially perpendicular to the disk. The outflow is 
stronger perpendicular to disk \citep{pro04} as the feedback is relatively
inefficient in the accreting disk of BH which supplies a continuous fuel to 
the BH via inflow. In addition, the massive and geometrically thick nature 
of the molecular torus \citep[e.g.,][]{kro88,tac94} may restrict the 
outflow to the direction essentially parallel to the disk angular momentum 
vector.

AGN outflows can collide with and shock ambient gas, generating a 
momentum-driven flow. To mimic this phenomenon, we let the emitted 
wind particle share its momentum with the neighboring gas particles. 
With this ``momentum share'' treatment, the two nearest neighboring gas 
particles are expelled together with the wind particle. They have the 
same velocity increment,  $\Delta \vw \sim 10,000/3~\kms$, conserving 
the momentum. Sharing momentum with other particles via inelastic 
collisions, however, decreases the total kinetic energy increment 
(Equation~(\ref{eq:edotw2})) while preserving momentum. We deposit the 
residual energy into these three particles in thermal form so that total 
energy is conserved. Note that the wind particles can reach very high 
temperatures. The analysis of the number of momentum sharing 
particles will be discussed in the later sections. 

Momentum sharing has two technical advantages. In general, high 
velocity particles will drive shocks. Similarly to SN remnants, the 
solution will approach a Sedov solution. However, with coarse mass 
and time resolution and having momentum share starts the cascade of
transforming the $\sim 100\%$ kinetic energy flux of the wind outflow to 
the $\sim 25 \%$ kinetic energy flux of the Sedov solution with twice the 
number of particles, a lower initial fraction of kinetic energy; and it 
makes the approach to the Sedov solution faster. Thus it makes us less 
sensitive to the problem of not having enough particles to correctly 
represent a hydrodynamic outflow. It also has computational advantages 
with regard to the time stepping. The standard Courant time step 
calculation implemented in the public release of GADGET might not 
guarantee fine enough time stepping for strong explosion problems 
\citep{spr10}, and requires additional time-step limiter implementation 
\citep{sai09,dur12}. In momentum sharing, reducing the velocity of the 
wind by a factor of three, while maintaining the same momentum flux, 
reduces the need for short and expensive time steps by the same factor, 
and our experiments have shown that if we perform solutions with high 
enough time and mass resolutions there is almost no difference 
between the two models. However, there is a great saving 
computationally in doing a momentum sharing when we are at low 
resolution.

In this approach mass, energy, and momentum are transferred to the 
ambient gas during AGN feedback and in addition to the usual 
efficiency parameter $\epsw$, we must introduce a parameter $\psi$
(Equation~(\ref{eq:psi})), the ratio of the outflow mass flux to the 
accreted mass flux, which is fixed by $\epsw$ and the wind velocity 
$\vw$.

\subsubsection{X-ray Feedback}
We also consider the radiative feedback from the 
electromagnetic energy component of X-ray radiation from the BH. 
We first calculate the total radiation emitted from the location of the 
BH particle as
\beq
L_{r} = \epsilon_{r}\Mdotbh c^2,
\label{eq:Lr}
\eeq
where the radiative efficiency $\epsilon_{r}=0.1$ is adopted in all 
simulations \citep{cio07,cio09,nov09}. This luminosity is converted to a 
luminosity flux at the position of each particle by $F_r = L_r / 4 \pi r^2$, 
where $r$ denotes the distance of the particle from the BH particle. We 
convert the flux to the net volume heating rate $\dot{E}$ by using the 
\citet{saz05} formulae, which describe the net heating rate per unit 
volume of a cosmic plasma in photoionization equilibrium with a 
radiation field characterized by the average quasar spectral energy 
distribution, as in \citet{cio10,nov11}. We take into account Compton 
heating and photoionization heating as summarized below. The 
volume heating rate $\dot{E}$ in cgs units is given as:
\begin{equation}
\dot E = n^2 (S_1 + S_2),
\end{equation}
where $n$ is the proton density (in number). The Compton heating 
term $S_1$ is
\begin{equation}
S_1 = 4.1\times 10^{-35} (1.9\times 10^7 -T)\,\xi,
\end{equation}
where the ionization parameter $\xi$ is defined as
\begin{equation}
\xi \equiv \frac{L_r}{n(r) r^2}.
\end{equation}
The sum of photoionization heating is given as
\begin{equation}
S_2 = 10^{-23}{a\, (\xi/\xi_0)^b\over 1 + (\xi/\xi_0)^b},
\end{equation}
where
\begin{equation}
a={1.7\times 10^4\over T^{0.7}}, 
\end{equation}
\begin{equation}
b=1.1-{1.1\over  e^{T/1.8\,10^5}}+{4\times 10^{15}\over T^4}, 
\end{equation}
and finally
\begin{eqnarray}
\xi_0 &=& {1\over 1.5/\sqrt{T}+1.5\times 10^{12}/\sqrt{T^5}}+\nonumber\\
      &&  {4\times 10^{10}\over T^2}
          \left[1 + {80\over e^{(T-10^4)/1.5\,10^3}}\right].
\end{eqnarray}

A speed-of-light delay in propagation from the BH is not included, since 
the effects of the delay should be small because of the small simulation 
scale ($\lesssim 50$ kpc). We neither include radiative transfer nor 
optical depth effects for the hard X-ray radiation considered here.

However, we do include the electromagnetic momentum---the radiation 
pressure from the X-ray flux from the BH as in \citet{deb10,deb10b}. We 
model the radiation pressure by applying a total momentum per unit time 
of
\beq
\dot{p}=\dot{E}/c,
\eeq
away from the BH particle to each gas particle, where $\dot{E}$ is the 
energy absorbed by the particle given the \citet{saz05} prescriptions. 
Here we neglect the effect of dust since the ISM generally has a low 
optical depth to hard X-rays. \citet{ham11} and \citet{kim11} recently 
studied the effects of X-ray radiation on the properties of massive 
elliptical galaxies. They found that X-ray feedback is more effective at 
suppressing star formation and BH mass growth compared to the 
traditional thermal feedback model.

\subsection{Eddington Force}\label{sec:eforce}
In the previous studies, the maximum accretion has been limited to the 
Eddington rate (Equation~(\ref{Eddington})) as shown in 
Section~\ref{sec:bhfeedback}. Instead of manually limiting the mass 
accretion, we actually compute the Eddington force (EF) acting on the 
surrounding gas particles, directed radially away from the SMBH. We 
first calculate the luminosity as in Equation~(\ref{eq:Lr}), and the flux at 
the position of the each particle by $F_r = L_r / 4 \pi r^2$ where 
$r$ denotes the distance of the particle from the BH particle. Then, the 
total momentum change per unit time by the EF acting on 
the gas particles radially away from the SMBH particle is given as
\beq
\dot{p} = \frac{F_r N_e \sigma_{T}}{c},
\eeq
where $N_e$ denotes the number of electrons associated with each 
gas particle and $\sigma_T$ is the Thomson cross-section for the 
electron. When the SMBH has vigorous mass accretion bursts, i.e., 
above the Eddington mass accretion limit, the strong radiation pressure 
by the SMBH pushes the gas particles away, resulting in a lower density 
near the SMBH particle, which leads to the lower mass accretion rate. 
Since the Thompson cross-section is independent of frequency ($h\nu \ll 
m_e c^2$), no radiative transfer treatment is required; we assume that 
absorbed UV radiation is re-radiated as IR radiation.

\subsection{New BH accretion model}
\label{sec:accretionmodel}
\subsubsection{Bondi Radius Criterion}\label{sec:bondi}
The key to understanding the accretion process lies in correctly 
modeling the behavior of the accreting gas once it falls within the 
gravitational influence of the BH, the Bondi radius, $\Rbondi$,
defined as
\beq
\Rbondi \equiv \frac{2G\Mbh}{{v_i}^2}\label{eq:rbondi},
\eeq
where ${v_i}^2=c^2 + {v_{\rm rel}}^2$, $c$ and $v_{\rm rel}$ denote
the sound speed and the relative velocity to the SMBH of the gas 
respectively. The Bondi radius divides the flow into two distinct regions. 
Far outside of $\Rbondi$, gas is hardly aware of the existence of the 
BH, and the flow is subsonic. On the other hand, inside the Bondi 
radius, gas has negative total energy and essentially plunges at free-fall. 
The Bondi radius in the spherical case being the place where the Mach 
number of the flow is unity. We note that heating that occurs within the 
Bondi radius cannot, by definition, affect the accretion rate since 
information cannot propagate upstream in a supersonic flow.

Applying the Bondi--Hoyle--Lyttleton formalism (Equation~(\ref{Bondi})) 
to the BH mass accretion rate assumes that the accretion onto the 
SMBH is commensurate with the accretion rate through the Bondi 
radius. In cases where the SPH smoothing length is greater than the 
Bondi radius, i.e., the inner flow is numerically unresolved, unusual 
effects can occur. To see what would happen if we used the standard 
accretion algorithm without feedback, we have made some artificial 
tests without any feedback mechanisms. We obtain a very large growth 
of the SMBH with $\Mbh \sim 10^{10} \Msun$. The neighboring gas 
particles around a BH at any distance from the BH are accreted, even for 
particles which are not gravitationally bound to the BH.

To avoid any unphysical accretion from outside of the Bondi radius,
we limit the accretion of mass to the gas particles statistically within the 
Bondi radius. With this Bondi radius criterion, gas particles can only be 
accreted when $r_i < \Rbondi $, where $r_i$ is the distance of the gas 
particle from the BH particle. When the mass of the BH is  small, 
we cannot resolve the Bondi radius, i.e., the smallest resolvable 
length scale of our simulations, the gravitational softening length, is of a 
few tens of pc, whereas the Bondi radius is just a few pc when the BH 
mass is around $10^{5}-10^{6} \Msun$ 
($\Rbondi$/pc = 3.4~$ M_{\rm BH,6}/v_{i,50}^2$ where
$M_{\rm BH,6}=10^6 \Msun$ and $v_{i,50}$=50~$\kms$).
In this case, we set the Bondi radius to be the smallest resolved scale, 
i.e., the gravitational softening length of the gas particles.

\begin{figure}
\epsfig{file=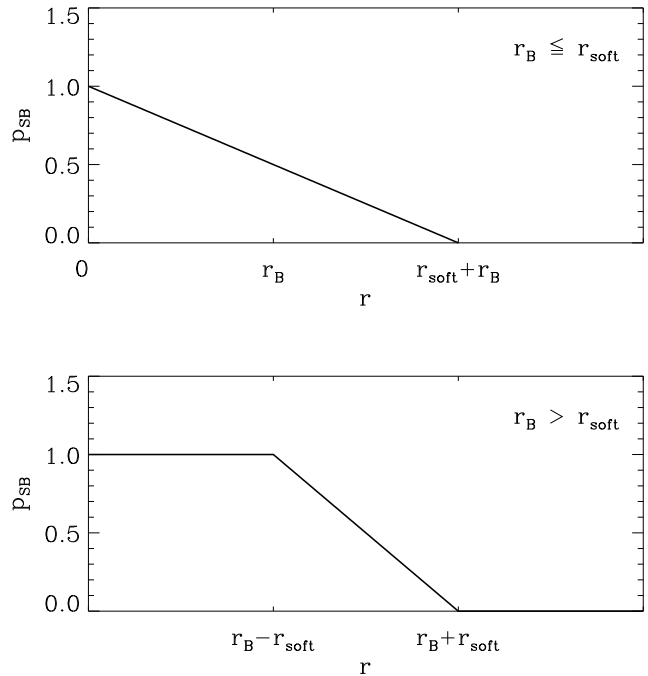,width=\columnwidth}
\caption{Soft Bondi probability of being absorbed by BH as a 
function of the particle distance $r$. The probability for the case when 
the Bondi radius $\Rbondi$ is equal to the softening length $r_{\rm soft}$ 
shown in the upper panel, and one for the case when $\Rbondi$ is 
larger than $r_{\rm soft}$ is shown below.
\label{fig:sb}}
\end{figure}

Since the gas mass distribution is smoothed with a kernel size $h_i$,
we apply what we term ``a soft cut of Bondi radius criterion''. We allow 
the full accretion rate, only when  the distance of the gas particle from 
the SMBH particles $r_i$ is smaller than $\Rbondi + r_{\rm soft}$. The 
gravitational softening length $r_{\rm soft}$, which avoids numerical 
singularities in the integral representation of the potential, is the 
smallest resolvable length scale and serves as the minimum bound to 
the smoothing length $h_i$. The soft Bondi probability $p_{\rm SB}$ as 
a function of the particle distance from the BH, for two cases of $r_B \le 
r_{\rm soft}$ and $r_B > r_{\rm soft}$, is shown in Figure~\ref{fig:sb}. 
When the soft Bondi limit (SB) is included, we reduce the probability of 
being absorbed by the BH for each gas particle (given as 
Equation~(\ref{p_acc})) used in the original code by the soft Bondi 
probability $p_{\rm SB}$, i.e., the final probability will be given as
$p_{\rm final}=p_j \times p_{\rm SB}$. This treatment essentially limits
the mass accretion only to the gas particles statistically within the 
Bondi radius.

\subsubsection{Free-fall modification of accretion rate}
In the standard version of the code the actual accretion of the gas
particles is determined by the probability, which is only a function of the 
SPH smoothing kernel. We have altered this to include a dependence 
on the time that it would take for the particle to be accreted allowing an 
extra factor of
\beq
p_{j,\rm ff} = \frac{ \frac{1}{\tau_j}}
{ \frac{1}{N_{\rm sph}} \sum\limits_{j=1}^{N_{\rm sph}} \frac{1}{\tau_j}},
\label{p_ff}
\eeq
where  $\tau_j = {r_j} / (c_{\rm s}^2+v^{2})^{1/2}$ is the free fall time and 
$N_{\rm sph}$ denotes the typical number of smoothing neighboring gas 
particles of the BH. We modify the probabilities (Equation~(\ref{p_acc})) to 
make them proportional to $p_{j,\rm  ff}$, i.e., we make it more likely 
that nearby particles will be accreted in a given time step than more 
distant ones. When the free-fall (FF) modification is included, the  final 
probabilities of the particles to be accreted by BH are given as 
$p_{\rm final}=p_j \times p_{j,\rm ff}$. where $p_j$ denotes a probability 
of being absorbed by BH used in the original code 
(Equation~(\ref{p_acc})).

\subsubsection{Alternative averaging in the mass accretion calculation}
In the original implementation of \citet{spr05}, the BH mass accretion is 
calculated based on the physical quantities such as density, sound 
speed, and relative velocity of the surrounding gas of the BH as shown 
in Equation~(\ref{Bondi}). These physical quantities for the BH particle are 
calculated from the $\sim 64$ neighboring gas particles using the SPH 
kernel. We can rewrite the Equation~(\ref{Bondi}) as,
\begin{equation}
\dot{M}_{\rm{B}}=
            \frac{4 \pi \alpha G^{2} M_{\rm BH}^{2} \langle \rho \rangle}
            {(\langle c_{\rm s} \rangle^2+\langle v \rangle^{2})^{3/2}},
\label{bondi_notAA}
\end{equation}
where angle brackets denote the averaging over the neighboring 
particles using SPH kernel. This method can be unstable to the number 
of SPH neighboring particles chosen and it separately averages 
quantities in the numerator and the denominator of 
Equation~(\ref{bondi_notAA}). Suppose we made a run which used the 128 
nearest particles instead of 64 nearest particles. It would have a 
different evolution, the soundspeed would be lower and so would the 
density. As noted earlier it is not the number of particles, per se that 
matters. If we halved the mass resolution and then moved to 32 
particles we would keep the total mass into which the feedback energy 
was deposited a constant. We have performed this experiment and find 
that results are essentially unchanged. However, of course putting the 
energy into increasingly more mass would lower the feedback induced 
increase in $c_{\rm s}^2$ and change the Bondi rate after the use of 
Equation~(\ref{Bondi}).

Our new averaging method for the calculation of the BH mass accretion 
``alternative averaging (AA)'' does the calculation in both time and 
space on an individual particle basis and then averages the results over 
the nearest 64 particles in order to reduce the dependency on the 
number of SPH particles. With AA, we can rewrite Equation~(\ref{Bondi}) as,
\begin{equation}
\dot{M}_{\rm{B,AA}}= 
\left\langle \frac{4 \pi \alpha G^{2} M_{\rm BH}^{2} \rho }
                            {(c_{\rm s}^2+ v^{2})^{3/2}} \right \rangle,
\label{bondi_AA}
\end{equation}
where angle brackets denote the averaging over the SPH kernel.
This method of performing the averaging is convergent because the 
added outer particles add progressively less and less.

\subsubsection{Fiducial BH mass accretion model}
In the fiducial model, we include all of the modifications we described for the 
BH mass accretion model, i.e., soft Bondi radius criterion (SB), 
free-fall modification (FF), and alternative averaging (AA). We first 
calculate the BH mass accretion rate using Equation~(\ref{bondi_AA}) as 
shown above, and for the actual accretion of the gas particles onto the 
BH particle, we calculate a probability of being absorbed for each gas 
particle $j$ as
\begin{equation}
p_{j,\rm final} = p_j \times p_{j,\rm SB} \times p_{j,\rm ff}.
\label{final_p}
\end{equation}

We test our modified BH accretion model against the analytic Bondi 
solution in an idealized homogenous environment. As demonstrated in 
the Appendix we recover the Bondi solution.

\begin{table*}
   \begin{center}
   \caption{Summary of the  Numerical Resolution}
   \vskip+0.5truecm	
    {
   \begin{tabular}{c|c|c|c|c|c|c|c}\hline\hline
Model & DM Particles & Disk Particles & $m_{\rm DM}$ ($h^{-1} \Msun$) & $m_{\rm bar}$ ($h^{-1} \Msun$) & $\epsilon_{\rm bar}$$^{\rm a}$ & $\epsilon_{\rm DM}$$^{\rm a}$ & $\alpha$$^{\rm b}$\\
  \hline
Low$^{\rm c}$  & 30,000 & 40,000 & $2.96 \times 10^7$& $3.91 \times 10^5$&0.1 & 0.8 & 100\cr
High & 400,000 & 300,000 & $2.25 \times 10^6$ & $1.30 \times 10^5$  & 0.02 & 0.083 & 35\cr
Very high & 800,000 & 600,000 & $1.13 \times 10^6$ & $6.50 \times 10^4$ & 0.016 & 0.066 & 32.5 \cr
Ultra high & 1,600,000 & 1,200,000 & $5.62 \times 10^5$ & $3.25 \times 10^4$ & 0.013 & 0.052 & 30\cr
  \hline\hline
   \end{tabular}}
   \end{center}
   \label{tab:res}
   \begin{flushleft}
    $^{\rm a}$ Gravitational softening lengths in $h^{-1}$ kpc.\\
    $^{\rm b}$ Dimensionless accretion parameter in Equation~(\ref{Bondi}).\\
    $^{\rm c}$ Numerical resolution used in \citet{spr05b} corresponds to our low-resolution model.
   \end{flushleft}  
 \end{table*}

\begin{table*}
   \begin{center}
   \caption{Summary of Model Properties\label{tab1}}
   \vskip+0.5truecm	
\resizebox{\textwidth} {!}{
  \begin{tabular}{c|c|c|c|c|c|c|c|c|c||ccccc}\hline\hline
  &   & \multicolumn{3}{|c|}{Feedback} &&&& & & log& log& log & log & \\ \cline{3-5}
& Model  &{\footnotesize T/M} & {\footnotesize X-Ray} & {\footnotesize X-Ray RP } & FF & AA & SB & EF & NEL & $\Delta\Mbh$ & $\Delta \Mwind$ & $l_{\rm BH}^{\rm eff}$$^{\rm a}$ & $L_{\rm wind}$$^{\rm b}$ & $v_{\rm wind}$$^{\rm c}$\\
{\footnotesize(1)} & {\footnotesize(2)} &{\footnotesize(3)}&{\footnotesize(4)}&{\footnotesize(5)}&{\footnotesize(6)}&{\footnotesize(7)}&{\footnotesize(8)}&{\footnotesize(9)}&{\footnotesize(10)}&{\footnotesize(11)}&{\footnotesize(12)} &{\footnotesize(13)} &{\footnotesize(14)} & {\footnotesize(15)}\\
  \hline
1 & Thermal$^{\rm d}$ & Thermal  &  &  &  &&& && 7.47 & 8.92 &  --2.80 &36.99 & 53.17\cr
2 & Fiducial$^{\rm e}$ & Momentum  & $\surd$ & $\surd$ & $\surd$ & $\surd$& $\surd$ & $\surd$ & $\surd$ & 7.38  & 8.24 &  --2.66 &40.72& 1828.1\cr
3 & Fiducial w/o XRP & Momentum  &  &  & $\surd$ &$\surd$&$\surd$&$\surd$ &$\surd$& 7.96 & 9.22 & --2.22  &41.37& 1287.9\cr
4 & Fiducial w/o EF  & Momentum &$\surd$ & $\surd$ &  $\surd$&$\surd$&$\surd$& &$\surd$&  7.41& 8.56 & --2.51  &41.07&1727.2\cr
5 & Fiducial w/o SB & Momentum  & $\surd$ & $\surd$ & $\surd$ & $\surd$& & $\surd$& $\surd$&  7.54 & 8.71 &  --2.37 &41.08&1743.9\cr

\hline
6 & N & No feedback  &  &  &  &&& &&  9.81 & \--&  --3.86 &\--&\--\cr
7 & N-SB & No feedback  &  &  &  && $\surd$& & & 8.89 & \--& --2.38  &\--&\--\cr 
  \hline\hline
   \end{tabular}}
   \end{center}
{\bf Notes.} Model names indicate the 
activated physics (symbol $\surd$) in the simulations as detailed in 
Column 3--10. For example, in Model ``Thermal'' only thermal energy 
feedback is allowed, while in Model ``N-SB'', no feedback is included, 
with the soft Bondi (SB) treatment.
\begin{flushleft}

$^{\rm a}$$l_{\rm BH}^{\rm eff} \equiv L_{\rm BH,opt}^{\rm eff} / L_{\rm Edd}$ 
           where $L_{\rm BH,opt}^{\rm eff}$ is the BH luminosity in the optical 
           band after absorption, i.e., as it will be seen from infinity. The 
           mean Eddington rates for last 1 Gyr are listed.\\
$^{\rm b}$The mechanical luminosity of the wind based on the wind mass 
            rate and wind velocity measured at 5 kpc from BH for last $\sim 1$ 
            Gyr. $L_{\rm wind} \equiv \dot{M}_{\rm wind} v_{\rm wind}^2 / 2$.\\
$^{\rm c}$Gas wind velocity at 5 kpc from BH in $\kms$.\\
$^{\rm d}$This model corresponds to the purely thermal feedback model
            discussed in SdMH05 \citep{spr05b}.\\
$^{\rm e}$Our best proposed ``Fiducial" model.\\
\end{flushleft}
\end{table*}

\section{Results}\label{result}
\subsection{Galaxy Models}\label{simulations}
The disk galaxies used in our study are set up in dynamical equilibrium 
and consist of a dark matter halo, a rotationally supported exponential 
disk of gas and stars, and a central bulge. The details of the model
construction are summarized in SdMH05. We test and study the stability 
of the constructed galaxy feedback models with a representative sample 
of our disk galaxy models in isolation with $v_{\rm vir} = 160~\kms$, 
and $r_{\rm vir} = 160 $ $h^{-1}$ kpc corresponding to a 
virial mass of $M_{\rm vir} = 9.53 \times 10^{11}$ $h^{-1} \Msun$. The 
dimensionless Hubble parameter is $h=0.71$ such that the present-day 
Hubble parameter is $H_0 = 71$ km $\rm s^{-1}$ $\rm Mpc^{-1}$. The 
\citet{her90} profile dark matter halos are constructed with the 
concentration parameter $c=9$ of the corresponding 
Navarro--Frenk--White (NRW) halo \citep{nav97}. The dark matter halo is 
then populated with exponential disks with a baryonic mass fraction of 
$m_d = 0.041$, so that a total disk mass $M_d = m_d M_{\rm vir}$ with a 
fractional gas content of $f_{\rm gas}=0.2$ with the rest being stars.

To study the effects of numerical resolution, we use four models with
different mass and spatial resolutions but with the same initial setup. 
The resolution details, including the number of particles, particle mass, 
and corresponding gravitational softening lengths are given in Table 1. 
The dimensionless accretion parameter in Equation~(\ref{Bondi}) was 
set to $\alpha=100$ in the previous studies \citep[SdMH05;][]{joh09}, 
which correspond to our low-resolution model. This value is quite a bit 
higher than the theoretical value of $\alpha \sim 1$. This discrepancy is 
due to the numerical resolution limits, that is, the gas density and sound 
speed at the location of the SMBH are estimated over the large--scale, 
resulting in artificially low densities and high sound speed. The higher 
value of $\alpha$ has been adopted empirically to correct for this 
resolution effect. For the model with the higher numerical resolution, we 
adopt the smaller value of the dimensionless accretion parameter that 
can result in the similar scale of early accretion history of low resolution 
model with $\alpha=100$. We run a series of simulations without any 
feedback prescription with the different values of $\alpha$ and adopt 
the $\alpha$ value that best reproduces the early accretion history of the 
low-resolution run with $\alpha=100$. We turn off all the BH feedback to 
remove the resolution dependency of the feedback prescription. The 
adopted values of $\alpha$ for each resolution are shown in Table 1. 

We summarize the properties of the models in Table~2. For reference, 
we first list the thermal feedback model ``T'', with the standard BH mass 
accretion and feedback model described in SdMH05. Note that model 
properties of standard model ``T'' in this paper are different from those of 
SdMH05, as we adopt ``Very High'' resolution as a standard resolution 
in this study, and the energy coupling efficiency $\epsw=5\times 10^{-4}$
which is 10 times smaller than the value adopted in SdMH05. We name 
our best proposed model with all the modifications that we have 
described ``Fiducial''. To isolate and compare the effects of each 
feedback model or modification, we include some simulations with one 
modification missing. We also include the ``No Feedback'' model: with no 
BH feedback in any form. Details of each model are shown in Table~2.

\subsection{ Comparison of the different feedback mechanisms}
\begin{figure}
\epsfig{file=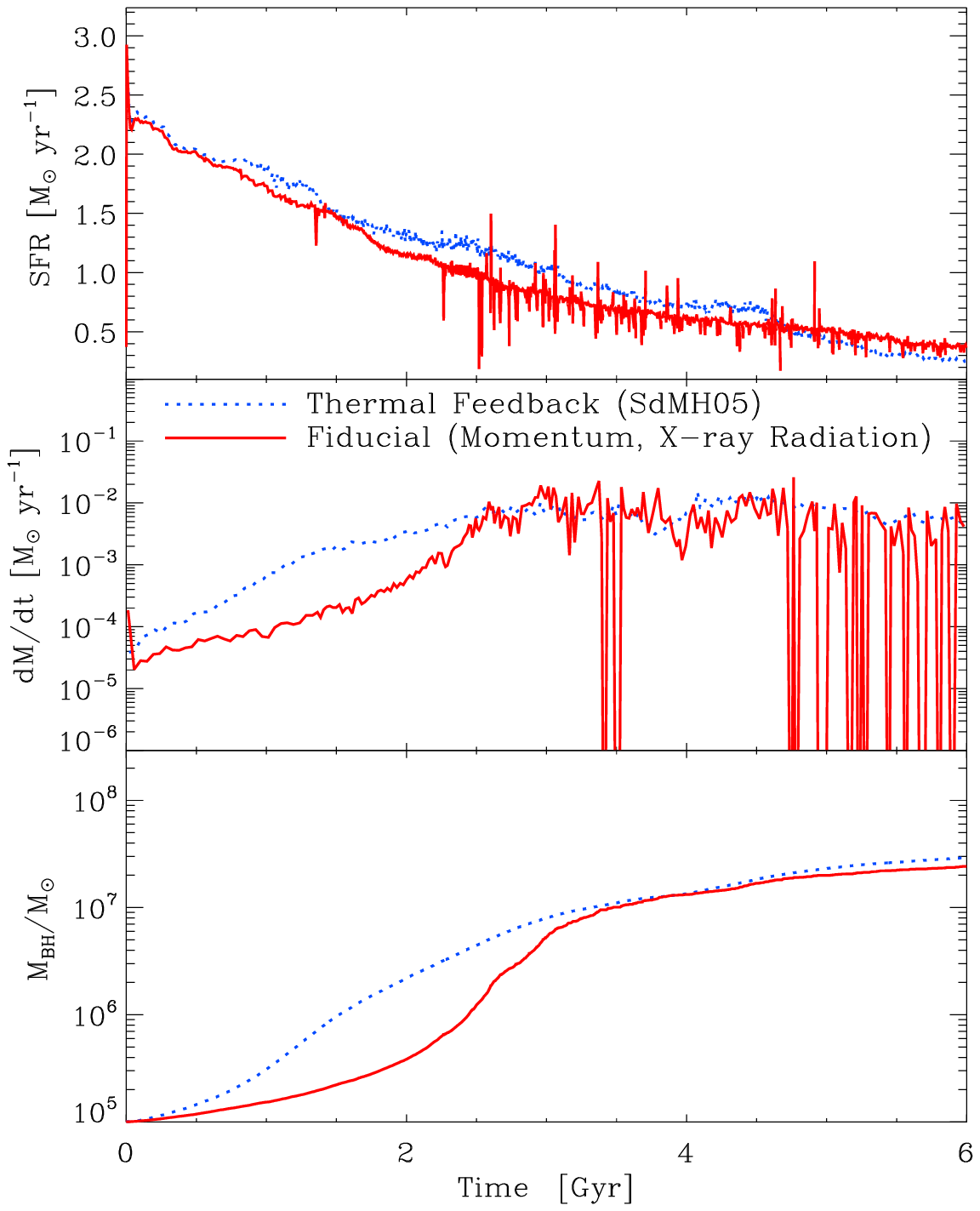,width=1.00\columnwidth}
\caption{Comparison of the feedback models: classical thermal 
feedback (blue), and our best proposed ``Fiducial'' model which has 
momentum mechanical feedback with the X-ray heating and radiation 
pressure (red). Evolution of the global star formation rate ({top}), the 
accretion rate onto the black hole ({middle}) and the evolution of the 
black hole mass ({bottom}) for an isolated gas-rich disk galaxy. Note 
that both the star formation rate and the black hole growth are essentially
similar in the new momentum driven treatment and the previous thermal 
feedback model. However, the degree of fluctuation in the mass 
accretion rate is greater in the momentum feedback model.
\label{fig:sfr}}
\end{figure}

We first examine the effects of various feedback mechanisms on the 
evolution of the BHs and galaxies. Figure~\ref{fig:sfr} shows the 
global star formation rate, the accretion rate onto the SMBH, and the 
evolution of the BH mass for the two different feedback models, 
the classical, thermal feedback (e.g., SdMH05), and our best 
proposed ``Fiducial'' model which has mechanical feedback with the 
X-ray heating and radiation pressure. Note that both models adopt the 
same feedback energy coupling efficiency $\epsw = 5\times 10^{-4}$, 
which is 10 times smaller than the value adopted in SdMH05. A seed 
SMBH starts at rest in the center with mass of $\Mbh=10^5 \Msun$ in 
all models, and grows due to gas accretion during the simulation. Note 
that the growth of BH is essentially the same in the two models. 
Thus the primary results of SdMH05 and \citet{dim05} with respect to 
the growth of central BHs during galaxy mergers are not 
substantially altered by the changes that we have introduced. Also, 
overall star formation rates of the two models are similar. However, the 
new model with mechanical feedback has more episodic star formation 
as in the mass accretion rate. Accretion rates and radiation output are 
much more variable in the new treatment with episodes of high 
accretion now reaching $\Lbh / \Ledd \geq 0.1$. The mechanical 
feedback model spends a large fraction of time at relatively low 
Eddington accretion rate, coinciding with the observed broad-line
active galaxies in the local universe \citep{gre07,ho09,kau09}.

Moreover, the two models have quite different wind properties. In the 
mechanical feedback model, wind particles are ejected with the 
instantaneous disk wind velocity of $\vw \sim 3000~\kms$ (OCCNP10), 
while the heating from AGN feedback energy drives slow and hot
outflows from galaxies in the thermal feedback model. The existence 
of a weak wind perpendicular to the plane of the disk in the vicinity of 
the BH was shown in SdMH05.

\begin{figure}
\epsfig{file=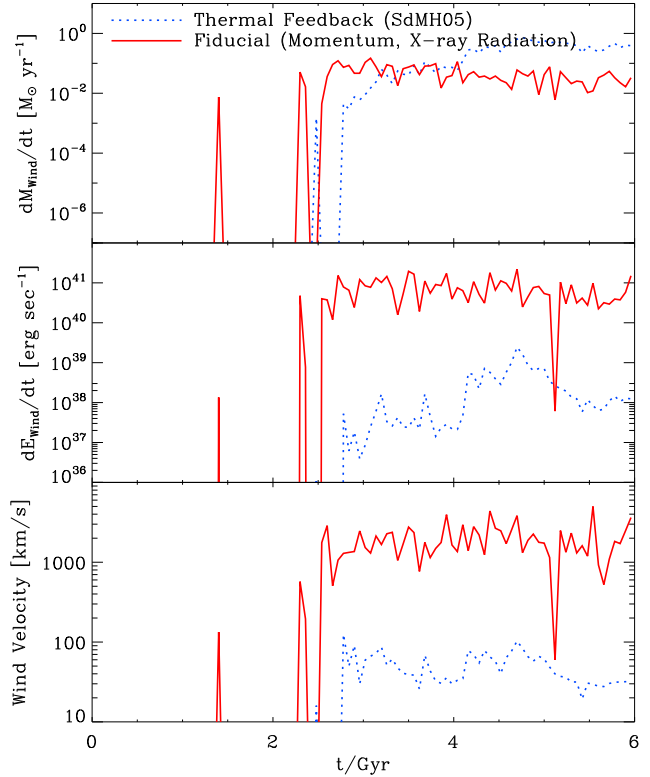,width=1.00\columnwidth}
\caption{Top: the evolution of wind mass loss rate; middle: 
corresponding mechanical luminosities; bottom: gas wind 
velocities. All quantities are measured at 5 kpc from the SMBH, above 
and below the disk midplane. Note that the thermal feedback model has 
much smaller wind velocity, a factor of 20 smaller compared to the 
mechanical feedback model.
\label{fig:wind}}
\end{figure}

In order to compare the wind properties, we first measure the total 
outflow wind mass throughout the simulation by summing up the mass 
of the gas particles which pass through the plane $|z|=5$ kpc, above 
and below the disk midplane. Then, we measure the velocity of the 
particle at each time step and calculate the corresponding mechanical 
luminosity as $L_{\rm wind} \equiv \dot{M}_{\rm wind} v_{\rm wind}^2 / 2$, 
i.e., the kinetic energy carried away by the outflowing winds. 
Temporal evolution of the wind properties (i.e., outflow rate, wind 
velocity and mechanical luminosity) of two representative models of
thermal feedback (SdMH05) and the fiducial model is shown in 
Figure~\ref{fig:wind}.  Once the BHs reach similar masses, after 
about 3 Gyrs, the thermal feedback model develops a wind with about 
a factor of 10 higher outflow rate, but the velocity of the wind 
$\vw \sim 100~\kms$, is a factor of 20 smaller  compared to the 
mechanical feedback model. The high wind velocity ($\vw \sim 2000~\kms$) 
in our fiducial model is consistent with the recent velocity 
measurement of mass outflows in local Seyferts \citep{fis11,mul11,
pou11} which ranges from $700~\kms$ up to $3000~\kms$.
 
The mechanical luminosities of the winds for two models are shown in 
the middle panel of Figure~\ref{fig:wind}, and the averaged mechanical 
luminosities for last 1 Gyr are listed in Table~2. The thermal feedback 
model has considerably smaller mechanical luminosity 
$L_{\rm wind} \sim 10^{37}~\ergs$, because of its slow wind velocity.
On the other hand, our fiducial model with mechanical feedback has 
strong outflow with $L_{\rm wind} \sim 10^{41}~\ergs$, which is 
consistent with the fact that large outflows with a kinetic power 
corresponding to a significant fraction of the AGN bolometric luminosity 
are commonly observed in X-ray observations of a number of quasars 
(mostly BAL systems) that reveal significant absorption 
\citep[e.g.,][]{cha03,cre03,pou03,hol08}. The total kinetic energy 
carried away by the winds, i.e., the mechanical luminosities 
integrated over the entire simulation time for the thermal model is
$\Delta \Ewind \sim 2.8 \times 10^{55}$~erg, while the momentum 
feedback model deposits $\Delta \Ewind \sim 8.3 \times 10^{57}$~erg
into the ISM within 6 Gyr.

From a feedback energy coupling efficiency $\epsw$ which is an input 
parameter, and a total mass change for the BH, we calculate the energy 
released by the BH for the two cases using $\Delta E_{\rm mech} = 
\epsw \Delta \Mbh c^2$ and compare the ratios of the energy released 
in winds to the energy released by the BH, 
$\Delta \Ewind / \Delta E_{\rm mech}$. While the thermal feedback 
model emits only 0.1 \% of the BH mechanical energy release in winds 
($\Delta \Ewind / \Delta E_{\rm mech}=0.0011$), the value rises to 38 \% 
for the momentum feedback model, $\Delta \Ewind / \Delta E_{\rm mech}
=0.38$. That is, the mechanical energy put into momentum drives winds
more efficiently than the energy put into heat.

\begin{figure}
\epsfig{file=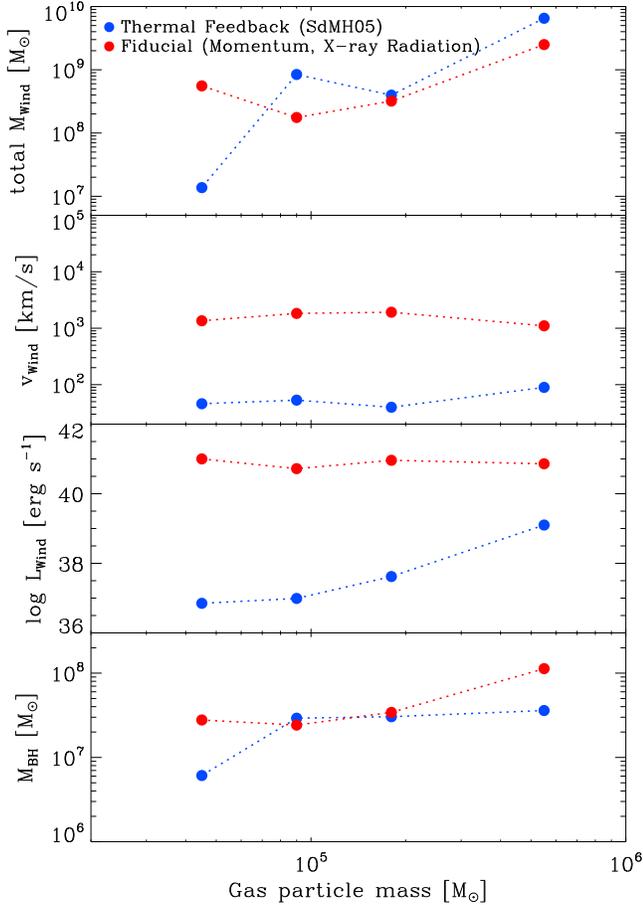,width=1.00\columnwidth}
\caption{Outflow wind properties and the final SMBH mass for the 
thermal feedback (blue) and the momentum feedback (red) are shown 
as a function of resolution (gas particle mass). The total outflow wind 
mass throughout the simulations, the averaged outward wind velocity, 
the averaged mechanical luminosity for last 1 Gyr, and the final SMBH 
mass (from top to bottom) are shown. In the thermal feedback 
model, much weaken winds are generated in the highest resolution, 
since the injected feedback energy is quickly radiated away due to the 
very short cooling time of the gas.
\label{fig:windres}}
\end{figure}

Now we investigate the behavior of the two models for disk galaxy 
simulations at higher numerical resolution. However, such a study is 
complicated because of numerical resolution issues. In the previous 
studies using one-dimensional simulations, it was found that including 
the mass and momentum component has the largest effect on the final 
mass of the SMBH, reducing the final SMBH mass by a factor of up to 
100 (OCCNP10). Turning on or off the energy input has relatively small 
effect, altering the SMBH mass only by a factor of two. But, in this 
three-dimensional study we have found that the momentum feedback is 
not more efficient than thermal feedback in protecting the SMBH from 
growth. However, the three-dimensional classic treatment has resolution 
effects that are difficult to correct because (1) the feedback energy is 
deposited outside of the Bondi radius whereas it is distributed within the 
Bondi radius in the much higher resolution one- and two- dimensional 
studies, and (2) the accretion is determined with the estimated gas 
density and sound speed averaged over the smoothing kernel, which is 
perhaps not the optimal procedure.

The results of the resolution dependency and the comparison of the wind 
properties of two feedback models are shown in Figure~\ref{fig:windres}.
Note that we adopt the smaller value of dimensionless accretion 
parameter $\alpha$ for the higher resolution runs as listed in Table 1. As 
described above, we measure the outflow wind properties throughout 
the simulations at $|z|=5$ kpc from BH, above and below the disk 
midplane. The measured wind masses, the time averaged wind 
velocities, and the averaged mechanical luminosities over the last 1 Gyr 
are  shown for each feedback model as a function of the gas particle 
mass in Figure~\ref{fig:windres}. The total amount of mass in the 
outflowing wind is resolution dependent in both models, but the effects
of resolution appear to be larger in the thermal feedback model. While 
the momentum feedback model in the highest resolution simulation has 
less wind mass (by a factor of 10) compared to the lowest resolution one, 
the wind mass difference between thermal feedback models at different 
resolutions reaches $10^3$. The wind velocity is resolution-independent 
for both feedback models, however thermal feedback models have much 
slower winds ($\sim 50-100~\kms$) compared to momentum feedback in 
all resolutions. The AGN-driven wind in the thermal feedback is also 
much slower than the velocity of the observed broad absorption line  
winds $\sim 10,000~\kms$ \citep{cre03,moe09}. In the case of the thermal 
feedback model at higher resolution, the shorter cooling time of dense 
gas makes thermal input increasingly inefficient at higher and higher 
numerical resolution. 

In the case of the final mass of the BH (bottom panel of 
Figure~\ref{fig:windres}), the effect of the resolution in the momentum 
feedback model is moderate, whereas the thermal feedback model has 
a factor of 10 smaller final mass in the highest resolution compared to 
one in the lowest resolution. For pure thermal feedback, we deposit the 
thermal feedback energy into the neighboring gas particles of the BH. 
The number of the affected neighboring gas particles is set to be the 
same for the different resolutions, therefore, in the higher resolution 
studies, we add energy into a smaller mass of the gas. In the higher 
resolution cases, gas particles in the central region have higher sound 
speed and lower density resulting in the lower BH accretion rate. 
However, the thermal feedback depends on the mass into which the 
energy is deposited, not on the number of particles. If the feedback 
energy were spread over constant mass by increasing the number of 
particles into which the thermal energy were added in the higher 
resolution study, the results would be essentially the same.

\begin{figure*}
\epsfig{file=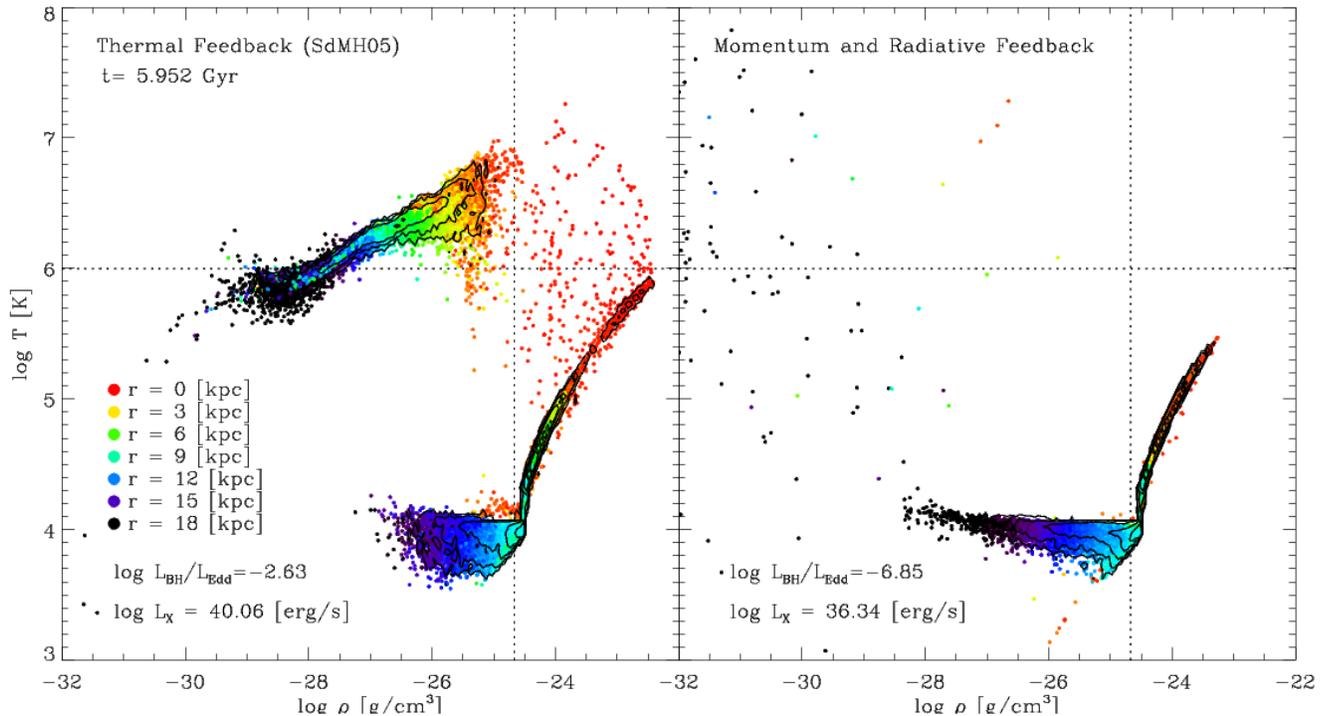,width=\textwidth}
\caption{Distribution of gas in the density--temperature plane at $t=5.95$ 
Gyr for the two feedback models. The radial distance from the BH is 
color coded, and the number density of SPH particles on the plot is 
shown with contours. The horizontal and vertical dotted lines 
respectively represent the temperature and density cuts we adopted for 
the X-ray luminosity estimate. Note that compared to the momentum 
feedback model, the thermal feedback model has much higher X-ray 
luminosity $L_{\rm X}  \sim 3 \times 10^{40}~\ergs$, which is emitted from 
the diffuse hot gas, even during the low BH accretion phase 
($L_{\rm BH} / L_{\rm Edd} \sim 0.002$).
\label{fig:rhoT}}
\end{figure*}

We compare the evolution of hot gas and X-ray emission of the two
feedback models. Following \citet{nav95}, we assume that X-rays are 
produced by the cooling of hot and diffuse gas, and estimate the X-ray 
luminosity for each SPH gas particle as,
\beq
L_{{\rm X},i}=1.2 \times 10^{-24} \frac{\rho_i m_{{\rm gas},i} }{(\mu m_p)^2} 
\left( \frac{kT_i}{1 {\rm keV}} \right)^{1/2} {~~\rm erg~s^{-1}},
\label{eq:xray}
\eeq
where $\rho_i$, $m_{{\rm gas},i} $ and $T_i$ are the density, mass, and
temperature of the {\it i} th gas particle in cgs units, respectively, $m_p$ 
is the proton mass, and $\mu$ is the mean molecular weight. The X-ray
emission computed via Equation~(\ref{eq:xray}) for each SPH particle is 
a lower limit as it assumes that the primary mechanism for X-ray 
emission is thermal bremsstrahlung. We do not include the X-ray 
emission via metal-line cooling although it is the more efficient cooling
mechanism for metal-enriched gas with a temperature of $\sim 10^6$ K 
and would produce more X-ray emission. We also assume that the 
central region of the galaxy remains obscured because of the large 
column density of intervening gas and dust, and calculate the total 
X-ray luminosity as
\beq
L_{\rm X,tot}=\sum\limits_{j=1}^{N_{\rm hot~gas}} L_{{\rm X},i},
\eeq
where the sum is over all hot and diffuse gas particles. Following 
\citet{cox06}, we define the `hot and diffuse gas' with a temperature of
$T \geq 10^6$ K, and a density $\rho \leq 3.16 \times 10^{-3} \Msun 
\rm pc^{-3}$, which corresponds to the critical density for star 
formation.

In Figure~\ref{fig:rhoT}, we show where the SPH particles lie in the
density-temperature plane with the calculated X-ray luminosities for 
both feedback models. In the thermal feedback model, the energy
input by the accreting BHs generates collimated, hot and slow 
winds perpendicular to the disk plane. These wind particles leave 
from the high density and hot temperature tip of the distribution of gas
in the density-temperature plane, and cool down slowly as they move 
outward. Due to their low velocity and high density, the cooling time is 
relatively long, and these hot and diffuse winds emit X-rays with the
luminosity far greater ( $\sim 10^{40}~\ergs$) than the momentum
driven winds ($\sim 10^{36}~\ergs$). The momentum feedback model 
has much lower X-ray luminosity as the momentum-driven winds 
quickly expel the X-ray emitting hot gas. Even during the low BH 
accretion phase with $L_{\rm BH} / L_{\rm Edd} \sim 0.002$, the 
X-ray luminosity of the thermal feedback model is much higher than 
what is typically seen from normal massive spiral galaxies at the 
present epoch ($10^{38}$--$10^{39}~\ergs$ in the $0.5$--$2.0$ keV band 
\citep{alm00,owe09,bor11}). The X-ray luminosity of the thermal 
feedback model is a lower bound, since the inclusion of 
cosmologically infalling gas in the simulations would lead to an 
increase in the computed thermal X-ray emission.

We now discuss the effect of the amount of the momentum-driven flow, 
i.e., number of momentum sharing neighbors in the momentum 
feedback model. As described in Section~\ref{sec:momentum}, we let 
the wind particle share its momentum with the neighboring gas particles, 
to mimic the shocked swept-up ambient medium. Because of the 
resolution limit, we keep the number of momentum sharing neighbors 
constant, despite the fact that a total swept-mass by momentum-driven 
wind depends on the mass of the BH \citep{kin03}. In order to study the 
effect of the number of neighbors to share the momentum feedback, we 
ran a series of simulations adopting different numbers of neighbors. 
{\it We find that the effects of sharing momentum with more gas particles 
are small.} Adding more particles is equivalent to assuming an early and 
brief transition to the Sedov phase. 

\subsection{Galaxy Merger Simulation}\label{merger}
\begin{figure}
\epsfig{file=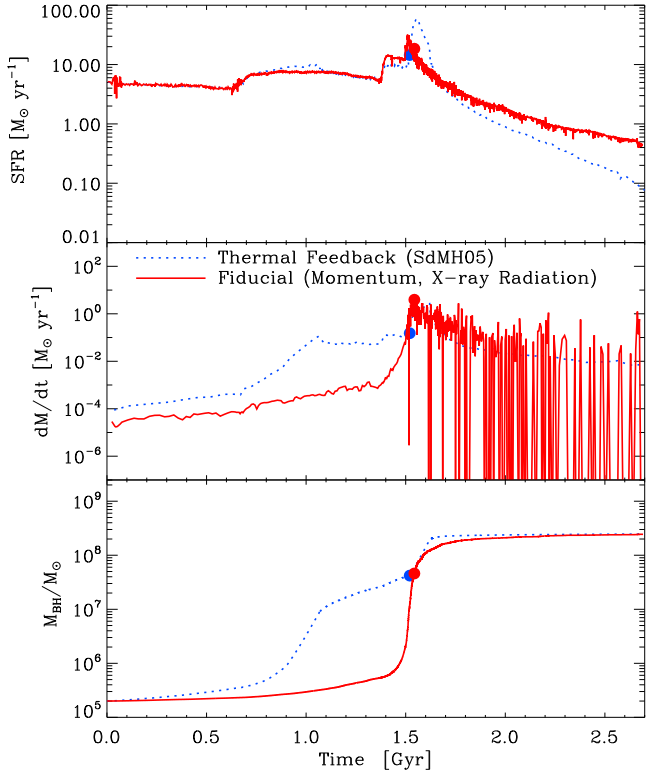,width=\columnwidth}
\caption{ Comparison of the feedback models with a major merger of 
two galaxies: classical thermal feedback (blue), and our best proposed 
``Fiducial'' model which has momentum mechanical feedback with the 
X-ray heating and radiation pressure (red). Evolution of the total star 
formation rate ({top}), the total accretion rate onto the black hole 
({middle}) and the evolution of the black hole mass ({bottom}) are 
shown as a function of time for the 1:1 merger. The filled circles 
indicate the time of SMBH merger. Note that both the star formation 
rate and the black hole growth are essentially similar in the previous 
thermal feedback model and the new momentum driven treatment.
However, the degree of fluctuation in the mass accretion rate is 
greater in the momentum feedback model as shown in the isolated 
galaxy case.
\label{fig:merger}}
\end{figure}

In addition to isolated galaxies, we also performed an equal-mass 
galaxy merger simulation using our progenitor disk galaxy models. The 
merger simulation was ran at high numerical resolution (see Table 1) 
with the initial seed BH masses set at $10^5 \Msun$. Following 
\citet{joh09b} we adopt  orbital geometry G13 \citep{naa03} for our 
merger simulation. This geometry corresponds to the inclinations 
$i_{p}=-109, i_{s}=180$ and the arguments of the pericenter 
$\omega_{p}=60, \omega_{s}=0$ for the primary and secondary 
galaxies, respectively. The galaxies approach each other on a 
parabolic orbit where the initial separation of the progenitors is 
$R_{\rm init}=r_{\rm vir}$ and the pericentric distance is 
$r_{\rm peri}=2 r_{\rm d}$, where $r_{\rm vir}= 160 h^{-1} \  \rm{kpc}$ is 
the virial radius and $r_{\rm d}= 2.5 h^{-1} \  \rm{kpc}$ is the disk scale 
radius. The simulation was evolved for a total of $t=3 \ \rm Gyr$ with the 
merger taking place at $t\sim 1.5 \ \rm Gyr$. The equal-mass merger 
initial conditions are simulated using both the standard thermal 
feedback and the new momentum mechanical feedback with the X-ray 
heating and radiation pressure.

In Figure~\ref{fig:merger}, we show the evolution of the resulting total 
star formation (top), SMBH accretion (middle), and SMBH mass (bottom) 
for the two feedback models as a function of time. A similar evolution is 
seen in the star formation rate for both models, however the new 
feedback fiducial model has episodic outbursts of mass accretion as 
shown in the isolated galaxy model. Even though the two feedback 
models have different SMBH accretion history, the final mass of black 
hole is essentially the same. Other aspects of the two simulations, such 
as the X-ray thermal luminosity, are significantly different, and this will be 
discussed in later papers.

\subsection{Soft Bondi criterion}
\begin{figure}
\epsfig{file=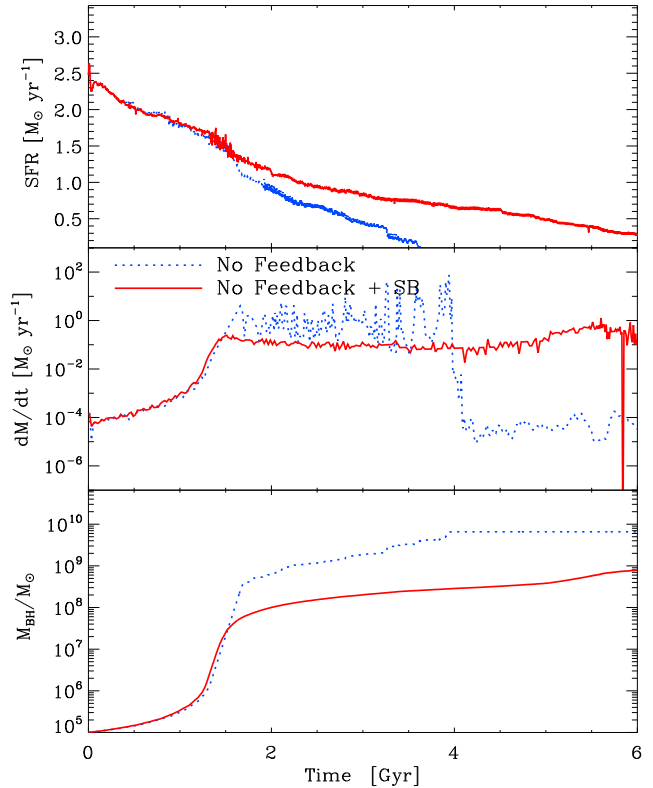,width=1.00\columnwidth}
\caption{Comparison of the models with and without the Bondi radius 
criterion. Evolution of the global star formation rate ({\it top}), the 
accretion rate onto the black hole ({\it middle}) and the evolution of 
the black hole mass ({\it bottom}) are shown for each model. Note the
order of magnitude reduction in the total accreted mass occasioned by 
including the Bondi limiter.
\label{fig:bondi}}
\end{figure}

Next, we turn our attention to the Bondi radius criterion. In order to test
the new soft Bondi mechanisms for limiting accretion when numerical
resolution is less than optimal, we ran two artificial test runs without 
any feedback, i.e., ``No Feedback'' models, with and without soft 
Bondi criterion. As discussed earlier in the Section~\ref{sec:bondi},
since any closest neighboring gas particles from the SMBH at any 
distance are considered as potentially accreting particles, gas particles 
keep accreting onto the SMBH and finally BH ends up swallowing all 
the gas in the host galaxy in the ``No Feedback'' model without soft 
Bondi criterion. 

With the Bondi radius criterion added, we limit the accretion only to the 
gas particles statistically within the Bondi radius and we can prohibit 
the gas particles which are not within the gravitational influence of the 
BH from accreting onto the SMBH. In more realistic simulations 
which included feedback, the effects of our Bondi limitation are far less 
significant.

In Figure~\ref{fig:bondi}, we show the resulting SFRs, BH accretion 
rates, and BH mass growth for these two models with and without the 
Bondi radius criterion. Utilizing the Bondi radius criterion effectively 
regulates the mass accretion, reducing the final mass by a factor of 10. 

\subsection{Fiducial model and the effects of other physics}
\begin{figure}
\epsfig{file=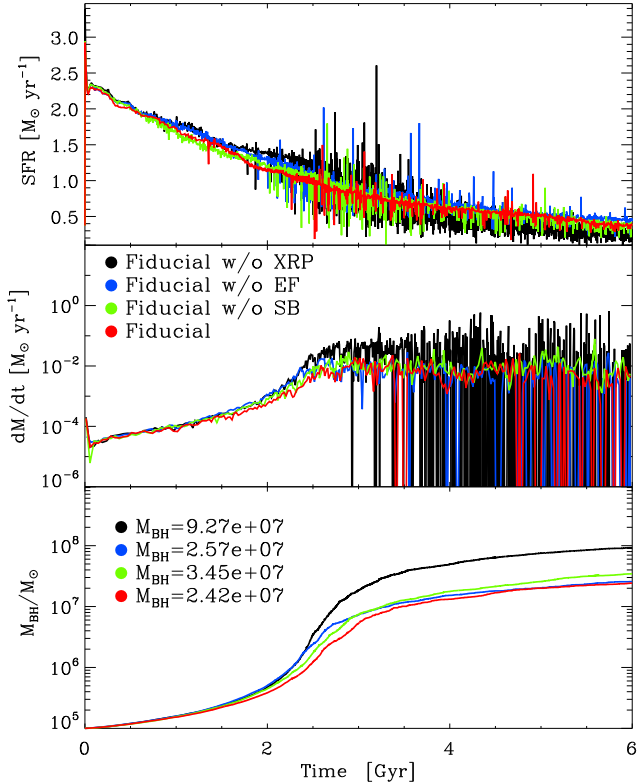,width=1.00\columnwidth}
\caption{Various model comparison. Evolution of the global star 
formation rate ({top}), the accretion rate onto the black hole 
({middle}) and the evolution of the black hole mass ({bottom}) in 
simulation of an isolated gas-rich galaxy. The largest of the newly 
included effects is clearly due to the X-ray heating and radiation 
pressure (XRP).
\label{fig:sfr_all}}
\end{figure}

In order to isolate and compare the effects of each feedback mechanism 
or modification made in the accretion, we perform the test simulations
with one modification missing. In Figure~\ref{fig:sfr_all}, we show the 
temporal evolution of SFR, mass accretion rate onto BH and BH mass of 
three distinctive models, the one without radiative heating and radiation 
pressure (fiducial w/o XRP), the one without the Eddington force 
(fiducial w/o EF), and the one without the soft Bondi radius criterion
(fiducial w/o SB), along with the fiducial model with all modifications as 
a reference. The radiative heating and radiation pressure is most 
effective among them, limiting the final mass of BH a factor of four. 
Compared to radiative feedback, both EF and Bondi 
criterion have minor effects on the final BH mass.

\section{Discussion}\label{discussion}
Absent a specific mechanism for transferring energy to particles 
surrounding an accreting BH, the AGN feedback via mass, and 
momentum output have not been included in classic thermal 
feedback works in three-dimensional hydrodynamic simulations. In 
this study, we have included the momentum mechanical feedback in 
the SPH simulation code, GADGET-2.
We also include a treatment of the feedback by the X-ray radiation 
which emanates from the BH and heats the surrounding gas 
in the host galaxies as well as radial momentum added to the fluid.
We statistically limit accretion to gas particles which are gravitationally 
bound to the central BH (Bondi radius criterion).

A series of test simulations of isolated systems with new 
implementations of the momentum feedback and radiative feedback 
with new criteria on mass accretion are performed. Relevant 
quantitative properties of the models are presented in Table~2, while 
the general results can be summarized as follows.

1. Overall, the BH growth is quite similar in the two 
approaches, so the successful prediction of the $\Mbh-\sigma$ relation 
by \citet{dim05,dim08} would be expected to be maintained in the new 
approach.

2. Our best proposed fiducial model with mechanical and radiative
feedback by hard X-rays has much higher velocity outflows compared 
to the thermal feedback model, with $\vw \sim 2000~\kms$ and 
$L_{\rm wind} \sim 10^{42}~\ergs$. The total emitted kinetic energy of 
mechanical feedback model is $\sim 100$ times higher than that of 
the thermal feedback model, even when the same feedback energy 
coupling efficiency is assumed and the BH growth is similar. The 
outflows found in our fiducial model have properties broadly similar to 
those observed in some local Seyfert galaxies \citep{ala11,fis11,mul11}.

3. The hot gas produced by slow, dense, thermally driven winds emits 
an X-ray luminosity significantly greater ($\sim 10^{40}~\ergs$) than the 
momentum-driven winds ($\sim 10^{36}~\ergs$). This X-ray luminosity of 
the thermal feedback model is also far greater than what is typically 
seen from normal spiral galaxies \citep[$\sim 10^{38}~\ergs$;][]{alm00,owe09}.

4. In the mechanical feedback model, the fluctuation level in both radiant 
and wind outputs is considerably greater than in the standard thermal 
feedback model.  While the thermal feedback model has a steady mass 
accretion with the Eddington ratio $L_{\rm BH}/L_{\rm Edd}=0.001$--0.01 
throughout the simulations, the momentum feedback model has 
stochastic bursts in the mass accretion with the Eddington ratio, which 
spans from $L_{\rm BH}/L_{\rm Edd}=10^{-6}$ to 1.

5. In an artificial model computed without any feedback mechanisms 
(no feedback model), the SMBH grows to $\sim 10^{10} \Msun$ 
accreting all the gas particles in the host galaxy. As noted, the 
standard prescription for accretion does not require the accreted 
particles be gravitationally bound to the central BH. However, the 
statistical implementation of the Bondi radius criterion can effectively 
limit the accretion of the gas particles to gravitationally bound particles, 
reducing the final mass of BH by a factor of 10. In more realistic models
with feedback the differences are small.

6. Radiative heating and radiation pressure on the ISM by photons
emitted by the central BH moderately reduces the final mass of BH,
by a factor of four.

7. The growth of the BH is confirmed to be essentially the same 
in the thermal and momentum feedback models in an equal-mass 
galaxy merger simulations.

\begin{acknowledgements}
We benefited from useful conversations with D. Clay Hambrick, Taysun 
Kimm, Gregory S. Novak, Daniel Proga, and Silvia Pellegrini. We thank 
MPA Garching for generously extending hospitality to the first two 
authors for very productive scientific visits. An understanding of the
standard approach was greatly aided by conversations with Lars
Hernquist and we gratefully acknowledge Volker Springel and Lars 
Hernquist for the use of the SPH code. We thank the referee for many 
suggestions and comments which greatly improved this manuscript.

J.P.O. and E.C. acknowledge the support of NSF Grant AST-0707505. 
T.N. and E.C. acknowledge the support from the DFG cluster of 
excellence ``Origin and Structure of the Universe''.  E.C. was supported
by the Samsung Scholarship foundation and made extensive use of the 
computing facilities of the Princeton Institute of Computational Science 
and engineering. TN acknowledges support from the DFG priority 
program SPP1177. P.H.J. acknowledges the support of Research 
Funds of the University of Helsinki.
\end{acknowledgements}

\appendix


\section{Accretion Model}\label{app}
We test our modified model of unresolved accretion onto the BH 
summarized in Section~\ref{sec:accretionmodel}. Our simulation set-up 
consists of a random distribution of gas particles in a periodic box with 
an accreting BH particle in the center. The size of the box is 1 kpc, and 
20000 gas particles with a mass of $2750~\Msun$ are used for the 
simulations. The mass of each gas particles is $\sim 30$ times smaller
than that of our galaxy simulation model at standard resolution, and the
gravitational softening length is $\sim 5$ pc. We choose $\Mbh=4 \times 
10^6 \Msun$ and the corresponding initial Bondi radius 
(Equation~(\ref{eq:rbondi})) is $\Rbondi = 10$ pc, with 
$T_{\rm gas}$=20000 K and Gaussian velocity distribution with 
$\bar{v}_{\rm gas}=50$ km/s and $\sigma_{v,\rm gas} = 5$ km/s. No 
radiative cooling is considered and dimensionless accretion parameter 
$\alpha$ is set to be 1.

\begin{figure}
\centering
\epsfig{file=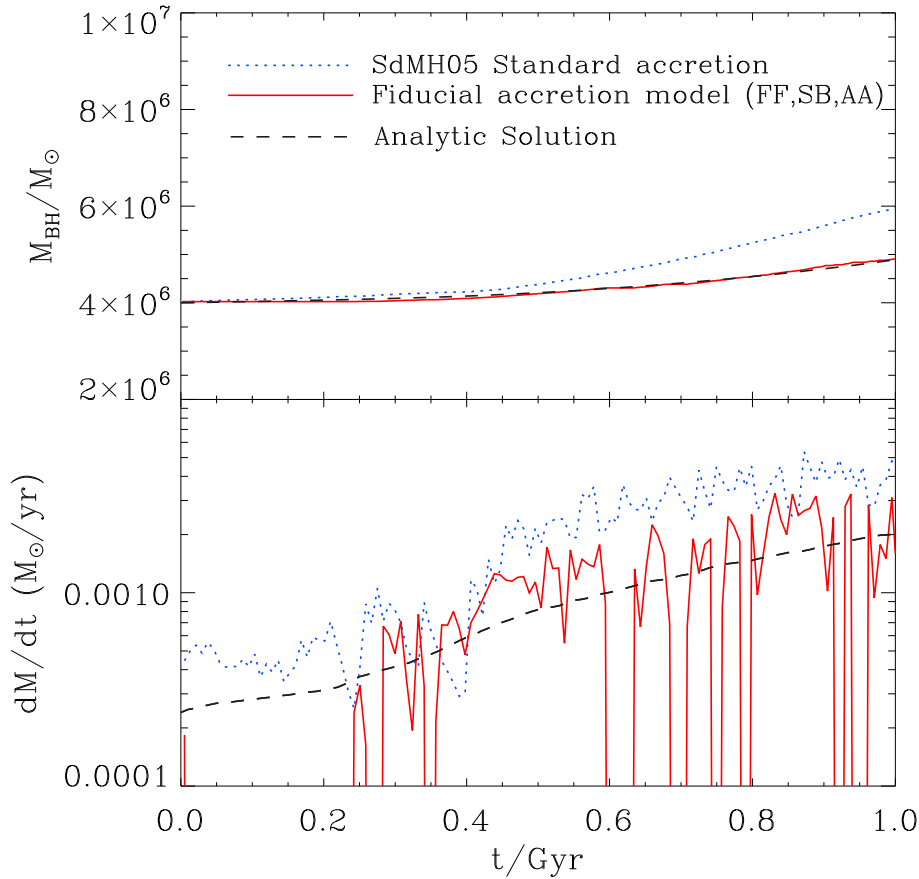,width=12cm}
\caption{Evolution of the black hole masses ({top}) and mass 
accretion rates onto BH ({bottom}) of the simulations of mass 
accretion models. The standard mass accretion model described in 
Section~\ref{sec:bhfeedback} in this paper (refer to SdMH05 for details), 
and new accretion model with all of modifications we described for the 
BH mass accretion model, i.e., soft Bondi radius criterion (SB), free-fall 
modification (FF), and alternative averaging (AA) (see 
Section~\ref{sec:accretionmodel}). The integrated new accretion model 
(red) agrees well with the analytic expectation despite the fact that the 
instantaneous accretion rate is often higher due to the intervals when 
the instantaneous rate is zero.
\label{fig:appendix}}
\end{figure}

We perform a direct comparison of our new accretion model with the 
SdMH05 model. In Figure~\ref{fig:appendix}, we plot the resulting BH 
accretion rate, and total BH mass for the simulation performed using the 
standard mass accretion model described in SdMH05, and compare it 
to the corresponding output of the accretion model presented in this 
paper (with FF, SB radius criterion, and AA following Equations~(\ref{bondi_AA})
and (\ref{final_p})). We also plot the analytic solution of the mass 
accretion of a given physical properties, i.e., Bondi solution which 
is described in the analytical formula of \citet{bon52} under the 
assumption of spherical symmetry and negligible angular momentum 
as,
\beq
\dot{M}_{\rm B} = \lambda 4 \pi \Rbondi^2 \rho_{\infty} c_{s,\infty},
\eeq
where the dimensionless parameter $\lambda$ depends only on the 
adiabatic index of the accreting gas (for details, see \citet{bon52,jan09}). 
For an assumed adiabatic index $\gamma=5/3$, $\lambda=0.25$ 
\citep{bon52}.

The evolution of the BH mass as a function of time for the new 
accretion model agrees well with that of the analytic Bondi solution. The 
total  accreted gas mass for the previously used accretion model is 
about two times larger. As shown in the bottom panels of 
Figure~\ref{fig:appendix}, in the standard mode the mass is accreted 
continuously whereas in our fiducial model we have discrete accretion 
events. That is mainly because the soft Bondi radius criterion prohibits 
the gas particles which are not within the gravitational influence of the 
BH from accreting onto the BH and stops the accretion before 
the gas particles are found statistically within the Bondi radius. The 
new model does not have mass accretion for the first $\sim 0.2$ Gyr,
before the accretion flow is formed around the BH in the center. Even 
if the Bondi radius is not fully resolved as in many of the applications, 
we obtain the Bondi solution with our new BH statistical mass accretion 
prescription. Given the discrete nature of our particles the accretion rate 
is inevitably stochastic. However, the computed accretion rate closely 
follows the exact analytic solution when we adopt our statistical 
treatment.

\end{document}